%% file: draft.tex
\documentclass[aps,a4paper,twocolumn,english,superscriptaddress,eqsecnum,longbibliography]{revtex4-1}

\usepackage{enumitem}
\usepackage{xcolor}
\usepackage{graphicx}
\usepackage{amssymb}
\usepackage{amsmath}

\usepackage[colorlinks,
            citecolor=blue,
            linkcolor=red,
            urlcolor=blue]{hyperref}

\usepackage[all]{hypcap}

\begin{document}

\title{NMR relaxation in the spin-1 Heisenberg chain}

\author{Sylvain Capponi}
    \affiliation{Laboratoire de Physique Th\'eorique, IRSAMC, Universit\'e de Toulouse, CNRS, UPS, France}
    \affiliation{Department of Physics, Boston University, 590 Commonwealth Avenue, Boston, Massachusetts 02215, USA}
\author{Maxime Dupont}
    \affiliation{Department of Physics, University of California, Berkeley, California 94720, USA}
    \affiliation{Materials Sciences Division, Lawrence Berkeley National Laboratory, Berkeley, California 94720, USA}
    \affiliation{Laboratoire de Physique Th\'eorique, IRSAMC, Universit\'e de Toulouse, CNRS, UPS, France}
    \affiliation{Department of Physics, Boston University, 590 Commonwealth Avenue, Boston, Massachusetts 02215, USA}
\author{Anders W. Sandvik}
    \affiliation{Department of Physics, Boston University, 590 Commonwealth Avenue, Boston, Massachusetts 02215, USA}
    \affiliation{Beijing National Laboratory for Condensed Matter Physics and Institute of Physics, Chinese Academy of Sciences, Beijing 100190, China}
\author{Pinaki Sengupta}
    \affiliation{Division of Physics and Applied Physics, School of Physical and Mathematical Sciences, Nanyang Technological University, Singapore 637371, Singapore}

\begin{abstract}
    We consider the isotropic $S=1$ Heisenberg chain with a finite Haldane gap $\Delta$ and use state-of-the-art numerical techniques to investigate its dynamical properties at finite temperature, focusing on the nuclear spin-lattice relaxation rate $1/T_1$ measured in nuclear magnetic resonance (NMR) experiments for instance. In particular, we analyze the contributions from modes with momenta close to $q\approx 0$ and $q\approx \pi$ as a function of temperature. At high-temperature we observe spin diffusion, while at low-temperature we argue that a simple activated behavior $1/T_1 \propto\exp(-\Delta/T)$ can be observed only at temperatures much smaller than the gap $\Delta$.
\end{abstract}

\maketitle

\section{Introduction}

Quantum magnets have long served as a testbed for discovering and understanding complex quantum many-body phenomena. The rapid advances in synthesis and characterization techniques accompanied by successful modeling and simulation using a range of diverse theoretical tools have constantly pushed the frontiers of strongly interacting systems. However, while experiments have probed both static and dynamic properties, theoretical approaches have primarily focused on static properties. This is due to the technical constraints of calculating dynamic properties. Yet understanding the nature of and the underlying mechanism behind the low-lying excitations is a key facet of complex many-body systems, and dynamical response functions constitute a major source of information. For example, analysis of inelastic neutron scattering data provides the most reliable insight into the ordering of spins in a quantum magnet. In this work, we calculate the dynamic structure factor for the $S=1$ Heisenberg antiferromagnetic (AF) chain to elucidate the mechanism of spin relaxation probed in nuclear magnetic resonance (NMR) experiments. Specifically, we shall investigate the nature of the spin relaxation in a gapped $S=1$ spin chain at finite temperatures.

The $S=1$ Heisenberg AF chain in one dimension is a paradigmatic example of a gapped system as conjectured by Haldane~\cite{haldane1983_bis,haldane1983} and verified numerically~\cite{Nightingale1986,Nomura1989,White1993,Todo2001,Nakano2009} and experimentally~\cite{Buyers1986,Renard2003}. Its hamiltonian is simply given by:
\begin{equation}
    \label{eq:H}
    {\cal H} = J \sum_{i} {\bf S}_i \cdot {\bf S}_{i+1},
\end{equation}
where $J$ is the antiferromagnetic nearest-neighbor exchange and the numerical value of the spin gap is $\Delta\simeq 0.41\,J$~\cite{Nightingale1986,Nomura1989,White1993,Todo2001,Nakano2009}.

Dynamical properties in spin systems can be probed using for instance inelastic neutron scattering (INS) to measure the dynamical spin structure factor $S(q,\omega)$ (see definition below). For very low-temperature, this spectral function is dominated by a single magnon branch, with a minimum at momentum $\pi$ where the peak width and position are weakly temperature dependent~\cite{jolicoeur1994,Damle1998,Syljuasen2008,Essler2008}. At intermediate temperatures (relative to the gap $\Delta$ since we will use $k_B=1$ to convert temperatures into energy scales), the competition between quantum and thermal fluctuations make the problem quite difficult, and recent numerical work has shown that intra-band magnon scattering can lead to additional features in the spectral function~\cite{Becker2017,lange2018}.

In NMR spectroscopy, the nuclear spin-lattice relaxation rate $1/T_1$ gives access to the local and dynamical spin correlation function (see definition below). Since the system is in a Haldane phase with a finite spin gap $\Delta$, it is rather natural to expect a simple activated law $1/T_1\propto\exp(-\Delta /T)$ at low temperature. Indeed, such activated behavior was recently observed numerically in gapped $S=1/2$ chains~\cite{dupont2016,coira2016}. However, for the  $S=1$ AF chain, there are some predictions based on the low-energy effective field-theory, namely the nonlinear $\sigma$ model.  In the large-$N$ approximation, the simple activated law above was found~\cite{jolicoeur1994}. In a refined similar calculation, Sagi and Affleck confirmed this result up to $\ln(T/\omega_0)$ corrections, $\omega_0 \ll J$ being the NMR frequency (we fix $\hbar=1$ for convenience), and also extended the result to finite magnetic field and other anisotropic cases~\cite{sagi1996}.

Using a semiclassical approach to the O(3) nonlinear $\sigma$ model (although being integrable, finite temperature correlations are hard to compute), Sachdev and Damle improved the previous result, by taking into account spin diffusion which occurs at long time~\cite{sachdev1997}. Their result is
\begin{equation}
    \label{eq:sachdev}
    \frac{1}{T_1} \propto \exp\left(-\frac{3}{2} \Delta/T\right),
\end{equation}
with a factor $3/2$ in the activated law. Nevertheless, this semiclassical prediction might not be correct in a full quantum mechanical solution of the O(3) $\sigma$ model.

Thanks to progress that has been made in computing dynamical properties for integrable models~\footnote{While the microscopic $S=1$ AF Heisenberg model is not integrable, the low-energy effective field theory is indeed integrable.}, Konik claimed to obtain ``\textit{exact low-temperature expansions of correlation functions}'' for Haldane chains~\cite{konik2003}. He was able to improve on previous results~\cite{sagi1996} by including higher order terms, but he still recovered purely ballistic transport and a simple activated behavior. There are some subtleties in all approaches when taking the long time limit~\cite{damle2005} or the zero-field limit~\cite{DeNardis2019} and it could be that integrability or not of the model changes results qualitatively~\cite{sirker2011,dupont2019}.

From an experimental point of view, the situation is also not so clear, as can be summarized from the following results, all obtained for Haldane materials. Various teams have tried to extract the activation energy measured in $1/T_1$ to compare it to the spin gap $\Delta$, i.e. measuring a $\gamma$ factor defined as follows:
\begin{equation}
    \frac{1}{T_1} \propto \exp\Bigl(-\gamma \Delta/T\Bigr).
\end{equation}
Early experiments on the famous Ni(C$_2$H$_8$N$_2$)$_2$NO$_2$ClO$_4$ (NENP) compound have led to $\gamma\simeq 1$~\cite{Gaveau1990}, but the applied magnetic field has a strong effect on the spin gap value~\cite{Fujiwara1992,Fujiwara1993}; later studies on Y$_2$BaNiO$_5$~\cite{shimizu1995} and AgVP$_2$S$_6$~\cite{takigawa1995,takigawa1996} concluded instead that $\gamma\simeq 1.2$ or $1.5$ with some uncertainty from the fitting window, the experimental error bars or the nucleus which is probed by NMR.

An unbiased numerical study of the full quantum one-dimensional model is called for, and we shall provide results in the following sections. The rest of the paper is organized as follows. In Sec.~\ref{sec:mod_def}, we present the theoretical models and provide useful definitions. Section~\ref{sec:results} describes very briefly the numerical techniques and thoroughly discuss the results. Finally, we summarize our conclusion in Sec.~\ref{sec:conclusion} and discuss implications and open issues.

\section{Model and definitions}\label{sec:mod_def}

The finite-temperature dynamic structure factor is defined through the K\"all\'en-Lehmann spectral representation:
\begin{eqnarray}
    S(q,\omega)=\frac{3\pi}{\mathcal{Z}(\beta)}\sum_{m,n}&&\mathrm{e}^{-\beta E_m}|\langle n| S^z_q|m\rangle|^2 \nonumber\\
    &&\times\delta\Bigl[\omega-\left(E_n-E_m\right)\Bigr],
    \label{eq:lehmann_dyn_str_fac}
\end{eqnarray}
where the sum is performed over the eigenstates of the Hamiltonian~\eqref{eq:H} with the partition function $\mathcal{Z}(\beta)=\mathrm{Tr}(\mathrm{e}^{-\beta\mathcal{H}})$ at inverse temperature $\beta=1/T$. The factor of $3$ comes from $\mathrm{SU}(2)$ symmetry. The momentum space spin operators~\footnote{Note that, in some references, Fourier transform definitions may slightly differ by using a factor $1/\sqrt{L}$ instead of $1/L$.} are related to those in real space via
\begin{equation}
    \mathbf{S}_q= \frac{1}{L}\sum_r \mathrm{e}^{-\mathrm{i}qr}\,\mathbf{S}_r,
\end{equation}
where for periodic boundary conditions (PBC) the discrete momenta are given by $q=2\pi n/L,$ $n=0,1,2,\ldots,L-1$. Note that thanks to SU(2) symmetry, we can measure equivalently diagonal ($S^z S^z$) or transverse ($S^\pm S^\mp$) correlations and symmetrize the data accordingly. In this paper, we focus on the local dynamical spin correlation $G_\mathrm{loc}(\omega)=\sum_q S(q,\omega)$:
\begin{eqnarray}
    \label{eq:Gloc}
    G_\mathrm{loc}(\omega) & = & 3 \,\mathrm{Re} \int_{0}^{+\infty} \, \mathrm{d}t\,\mathrm{e}^{-i\omega t} \langle S_i^z(t) S_i^z(0)\rangle, \\
    G_\mathrm{loc}(\tau) & =  & \frac{3}{\pi}\int_{-\infty}^{+\infty} \mathrm{d}\omega\, G_\mathrm{loc}(\omega) \exp(-\tau \omega),
\end{eqnarray}
where we have a sum-rule $G_\mathrm{loc}(t=\tau=0)=\langle\mathbf{S}_i^2\rangle=2$ and a symmetry property: $G_\mathrm{loc}(-\omega)=\exp(-\beta\omega) G_\mathrm{loc}(\omega)$.

Quite interestingly, assuming a local hyperfine coupling between the nuclear and electronic spin (and setting this amplitude to one for simplicity), the $1/T_1$ nuclear relaxation rate probed by NMR can be obtained directly from:
\begin{equation}
    \label{eq:T1}
    \frac{1}{T_1} = G_\mathrm{loc}(\omega_0) = \sum_q S(q,\omega_0),
\end{equation}
where $\omega_0~(\ll J,~\ll T)$ is the NMR frequency, which in many cases can be taken to be zero in practice, although in some regimes such as spin diffusion, data do explicitly depend on the choice of $\omega_0$, and this issue will be of concern to us here. Generically, spin diffusion is expected to occur at high temperature~\cite{muller1988}, and corresponds to a regime where $1/T_1$ depends on the experimental parameters such as the NMR frequency $\omega_0$ or the applied magnetic field. Numerically, it will translate into an explicit dependence on the chosen cutoff (finite-time, non-zero frequency $\omega_0$ or finite system length).

\section{Numerical results}\label{sec:results}

On periodic chains of small length, up to $L=12$ typically, it is straightforward to perform exact diagonalization of the Hamiltonian~\eqref{eq:H} in order to get the full spectrum, using symmetries such as fixing $S^z_\mathrm{tot}$ or the momentum. Since we can also compute all matrix elements of observables such as $S^z(q)$, one can easily compute $S(q,\omega)$, and hence $G_\mathrm{loc}(\omega)$ using Eq.~(\ref{eq:lehmann_dyn_str_fac}). Since it is given as a sum of delta peaks, we have chosen to represent these spectral functions using histograms with a suitably adapted bin width.

The quantum Monte-Carlo (QMC) simulations were performed using the stochastic series expansion (SSE) method, which is based on importance sampling of the Taylor expansion of the operator $\mathrm{e}^{-\beta\mathcal{H}}$, see e.g. Ref.~\onlinecite{sandvik2010}. The analytic continuation is performed using a recently improved variant of an approach called stochastic analytic continuation (SAC)~\cite{Sandvik1998,Beach2004,Sandvik2016} which has been applied to various magnetic systems for instance~\cite{Qin2017,Shao2017,Shu2018}. The spectrum is represented by a large number of equal-amplitude delta peaks whose positions are sampled at a fictitious temperature $\Theta$ adapted to provide a good fit (in a $\chi^2$ sense) of the imaginary-time data while avoiding
overfitting.

As discussed in Ref.~\onlinecite{starykh1997} for the $S=1/2$ Heisenberg chain, it is better to first perform analytic continuation for each $q$ in order to obtain $S(q,\omega)$ and then sum to get $G_\mathrm{loc}(\omega)$ --- and the NMR relaxation rate $1/T_1$ --- rather than the converse. Indeed, there is more structure in Fourier space for $S(q,\omega)$. This is particularly important to reproduce spin diffusion at high temperature. As a result, at very low temperature for finite $L$, very sharp oscillations will appear since in this limit $S(q,\omega) \approx a_q \delta(\omega-\omega_q)$ is dominated by single gapped triplet excitation (triplon)~\cite{Takahashi1989} with (approximately) relativistic dispersion relation $\omega_q=\sqrt{\Delta^2+v^2 (q-\pi)^2}$~\cite{White2008} where $v$ is the spin velocity.

The QMC-SAC approach is supplemented by Matrix Product States (MPS)~\cite{schollwock2011} calculations that were  performed~\footnote{Our calculations are based on the ITensor C++ library, available at \url{http://itensor.org}.} on open chains of size $L=64$ with a maximum bond dimension of size $m=1000$. Unlike the SSE QMC, the MPS method is a ground state technique. The finite-temperature was simulated using a purification method, artificially enlarging the Hilbert space size or, equivalently, doubling the system size ($L=128$) to represent the mixed state as a pure state~\cite{verstraete2004}. Ultimately, within this approach, the system with which we work looks like a two-leg ladder where each leg holds physical or auxiliary degrees of freedom. The infinite temperature state corresponds to a tensor product of maximally entangled rungs, readily encoded at the beginning and time-evolved with $\mathrm{e}^{-\beta\mathcal{H}/2}$ to obtain the corresponding state at inverse temperature $\beta$. The imaginary-time evolution is performed using the time-evolving block decimation (TEBD) algorithm~\cite{vidal2004} along with a fourth order Trotter decomposition~\cite{hatano2005} (time-step $\tau=0.1$), where $\mathcal{H}$ only acts on physical degrees of freedom. When the desired finite temperature state is obtained, a real-time evolution by $\mathrm{e}^{-i\mathcal{H}t}$ is carried out using the same TEBD algorithm as for the imaginary-time evolution in order to obtain the local dynamical correlation function $\langle S^z_i(t)S^z_i(0)\rangle$~\cite{binder2015} that one can relate to the dynamical spin correlation $G_\mathrm{loc}(\omega)$ by a standard Fourier transform, see Eq.~\eqref{eq:Gloc}. To avoid finite-size effects, we measured the correlation in the middle of the chain, $i=L/2$. Note that the real time evolution of a quantum state produces a rapid growth of entanglement entropy~\cite{laflorencie2016} while the efficiency of the MPS formalism relies on low-entangled states through the area law. This drastically limits in practice the maximum time $t$ one can reach in a simulation. In order to push the limit further, we have used the trick which consists of evolving the auxiliary degrees of freedom with $-\mathcal{H}$ in real time~\cite{karrasch2013,Hauschild2018}. It is worth mentioning that we have tried to compute time-dependent correlation functions in momentum space with MPS, but finite size effects, especially open boundary effects led to a strong low-energy ($\omega\rightarrow 0$) contribution owing to the edge states of the $S=1$ chain, as observed in Ref.~\onlinecite{Becker2017}.

\begin{figure}[!t]
    \includegraphics[width=\linewidth,clip]{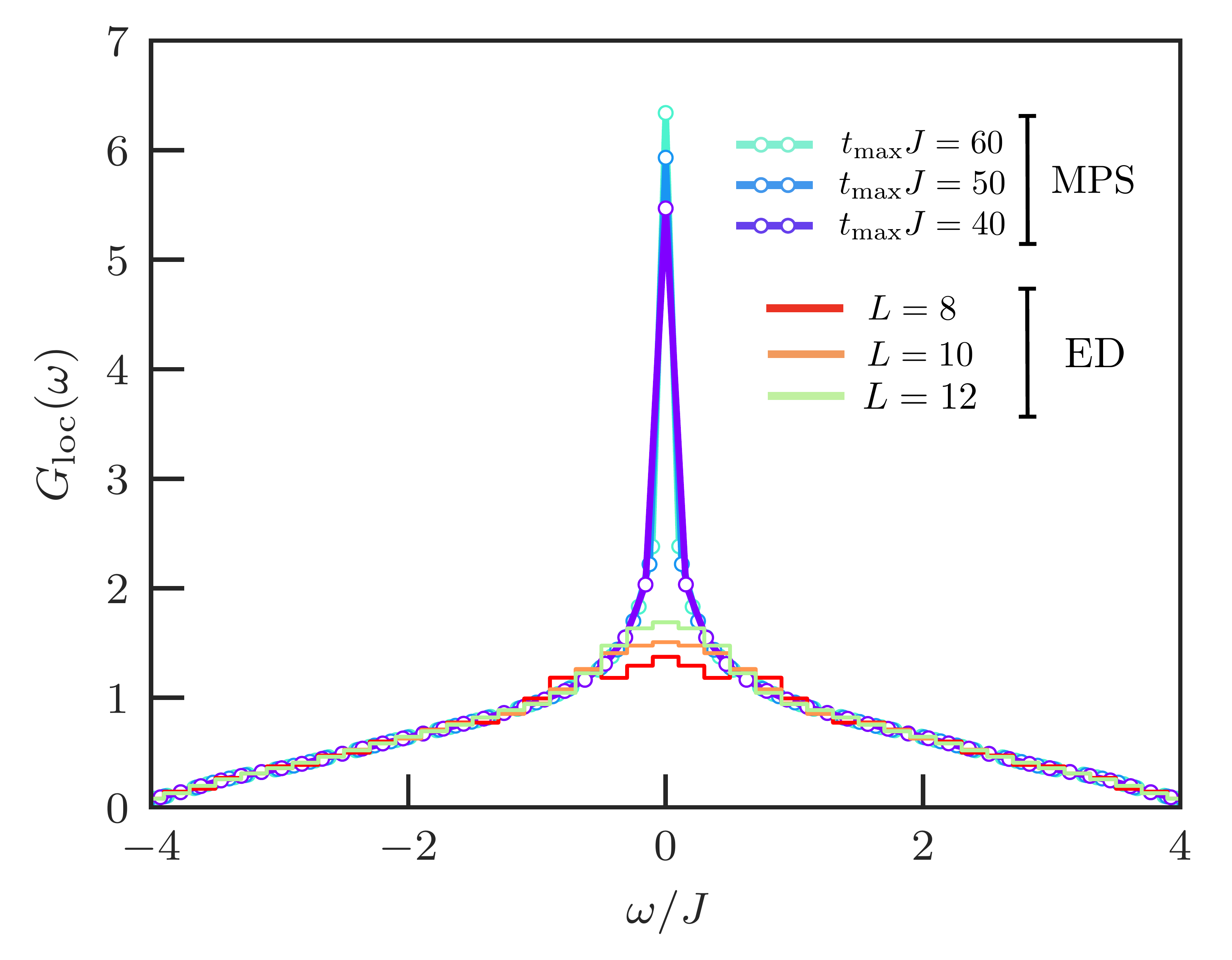}
    \caption{Comparison of the dynamical local spin-spin correlation $G_\mathrm{loc}(\omega)$ defined in Eq.~\eqref{eq:Gloc} at infinite temperature ($\beta J=0$) obtained from ED ($L=8$, $10$, and $12$) and MPS ($L=64$).}
    \label{fig:compare_beta0}
\end{figure}

\begin{figure}[!t]
    \includegraphics[width=\linewidth,clip]{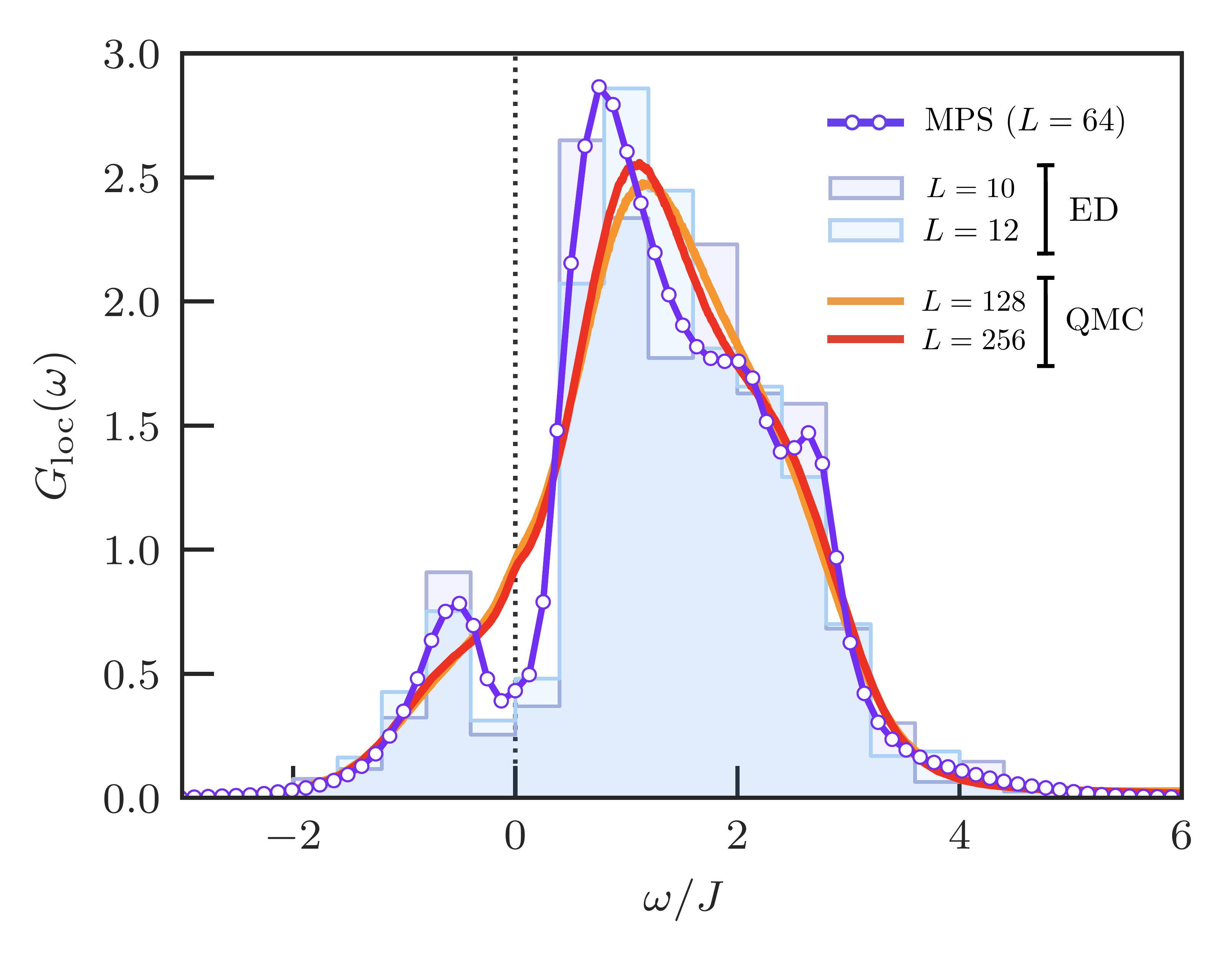}
    \caption{Comparison of the dynamical local spin-spin correlation $G_\mathrm{loc}(\omega)$ defined in Eq.~\eqref{eq:Gloc} at $\beta J=2$ obtained with different numerical techniques. Exact diagonalization (ED) was carried for systems of size $L=10, 12$. For the MPS, we used $L=64$ where the local dynamical correlation was measured up to time $tJ=50$ before performing the Fourier transform to frequency space. We also show QMC data after performing analytic continuation for $L=128$ and $L=256$. The different results and methods are discussed in the main text.}
    \label{fig:compare_beta2}
\end{figure}

In order to provide some benchmark for these numerical techniques, we plot in Fig.~\ref{fig:compare_beta0} the local spectral function $G_\mathrm{loc}(\omega)$ at infinite temperature ($\beta J=0$) obtained from ED or MPS. Quite nicely, the results look very similar for large $|\omega|/J$ although they are computed differently. On the other hand, the value at $\omega=0$ is not well defined (although it is a local quantity), which directly reflects the spin diffusion phenomenon: in ED, the numerical value depends on the length $L$; in MPS, the value explicitly depends on the real-time cutoff $t_\mathrm{max}$ limiting the simulation.

At an intermediate temperature, $\beta J=2$, we compare  the local spectral function $G_\mathrm{loc}(\omega)$ obtained with all three numerical approaches (Fig.~\ref{fig:compare_beta2}). Overall the agreement is quite good, although the QMC+SAC data cannot resolve the sharp peaks. It is known that at high-temperature, the small range of imaginary-time data $\tau \in [0,\beta/2]$ limits the accuracy of SAC~\cite{Shu2018}. Here it should be noted that both the ED and MPS spectra may still underestimate the spin diffusion contributions at the inverse temperature, $\beta=2$, used in the figure, but the QMC-SAC is most likely overestimating the spectral weight at low frequencies. Unfortunately, it is difficult to compare data at much lower temperature since MPS cannot be applied anymore accurately and ED becomes very sensitive to the discrete nature of a finite-size spectrum with small $L$. Nevertheless, the QMC+SAC has been proven in previous studies to provide reliable results at lower temperature~\cite{Sandvik2016,Qin2017,Shao2017,Shu2018}.

\subsection{Spin diffusion at high temperature}

\begin{figure}[t]
    \includegraphics[width=\linewidth,clip]{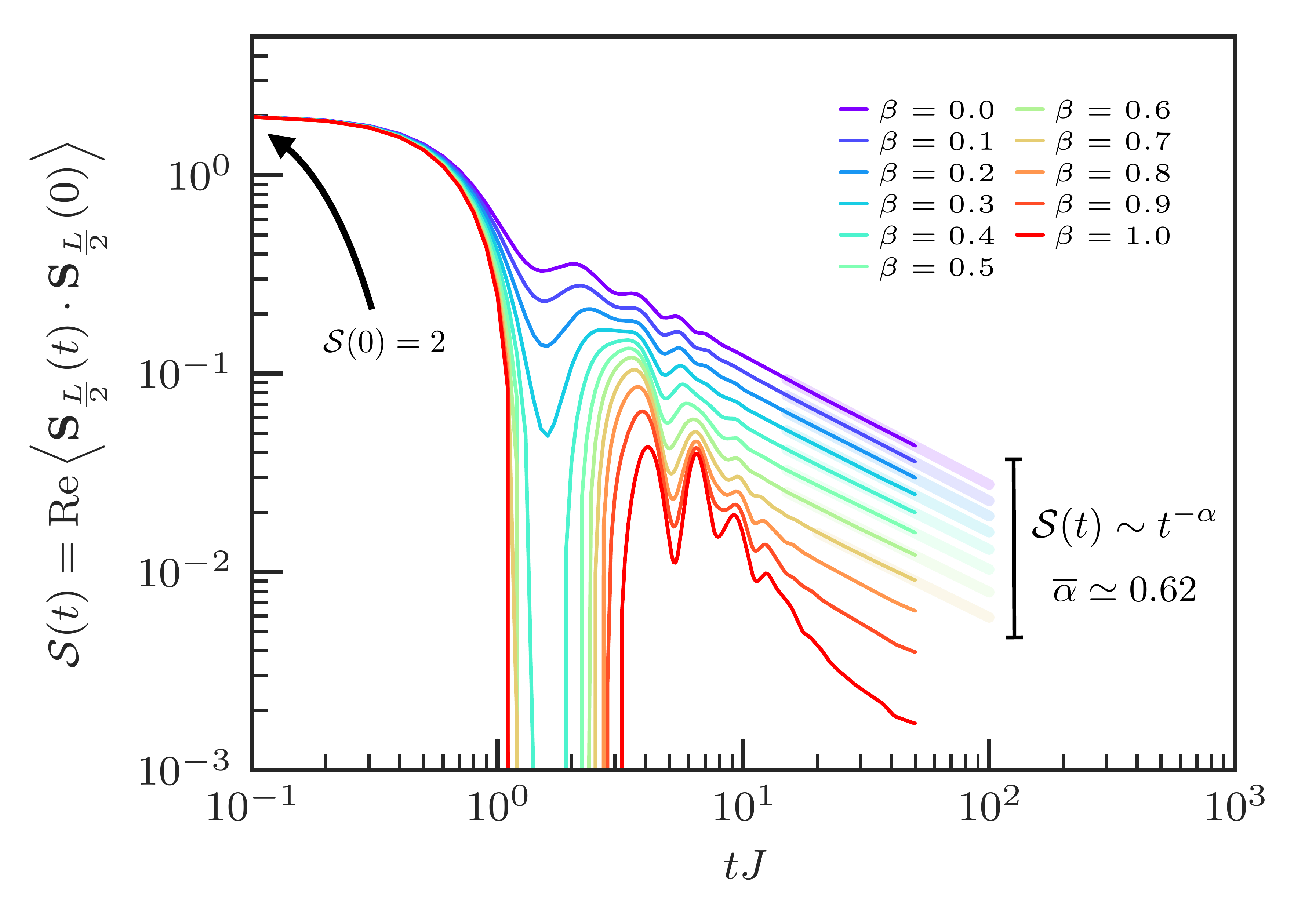}
    \caption{Real part of the local correlation function $\langle\mathbf{S}_{L/2}(t)\cdot\mathbf{S}_{L/2}(0)\rangle$ versus time for various inverse temperatures $0\leq\beta J\leq 1$, computed using matrix product states on $L=64$ spin chains. For $\beta J\lesssim 0.8$, a spin diffusion behavior is found, with an algebraic long-time decay $\propto t^{-\alpha}$ of the correlation function (see thick translucent straight lines in the log-log scale). An exponent $\overline{\alpha}\simeq 0.62$ independent of temperature is found.}
    \label{fig:spin_diffusion}
\end{figure}

On general grounds, a system is said to display spin diffusion if the long-time decay form of some spin correlation function is algebraic,
\begin{equation}
    \lim_{t\to\infty} \Bigl|\langle S^a_r(t)S^b_0(0)\rangle\Bigr| \propto t^{-\alpha_r},\quad\alpha_r\geq 0,
\end{equation}
where \textit{a priori} the exponent $\alpha_r$ could depend on the distance $r$~\cite{sirker2006}. This behavior was first predicted in a classical phenomenological diffusion theory describing the time-dependent spin correlation function of the classical Heisenberg model (the same Hamiltonian as in Eq.~\eqref{eq:H}, but where spin operators are replaced by unit length vectors)~\cite{muller1988,gerling1989,muller1989b,liu1991,alcantara1992,bohm1993,lovesey1994,lovesey1994b,srivastava1994}. This description is particularly valid at high temperature where quantum effects are suppressed and implies a conserved quantity (such as the total magnetization $S^z_\mathrm{tot}$ along the quantization axis) due the associated continuity equation. Specifically, a purely diffusive exponent $\alpha=d/2$ with $d$ the dimensionality is found for the classical Heisenberg model, which was verified numerically~\cite{bagchi2013}.

The next question is whether or not the exact treatment of the microscopic quantum model is consistent with this prediction. In this respect, the paradigmatic XXZ spin-$1/2$ chain parametrized by an uniaxial anisotropy $\Delta$ in the interaction has been intensively studied ~\cite{fabricius1997,starykh1997,fabricius1998,sirker2006,prosen2011,karrasch2013b,ilievski2015,bertini2016,bulchandani2018,Gopalakrishnan2019,gopalakrishnan2019b,DeNardis2019a}. The diffusion exponent $\alpha$ of the local correlation function $\langle S^z_0(t)S^z_0(0)\rangle$ along the quantization axis $z$ is exactly equal to one at the non-interacting point, as well as for the whole range $-1<\Delta<+1$, leading to ballistic spin transport. For $|\Delta|>1$, the bahavior of the spin dynamics has been found to be diffusive with $\alpha=1/2$, while more recently, $\alpha=2/3$, as in the Kardar-Parisi-Zhang (KPZ) universality class~\cite{Kardar1986} was observed at the isotropic $\Delta=1$ point~\cite{znidaric2011,ljubotina2019,Gopalakrishnan2019,gopalakrishnan2019b}. This is very different from the exclusively diffusive behavior that one gets with the classical Heisenberg model, but is now understood to be a consequence of the integrability of the quantum spin-half chain (the classical Heisenberg model is not integrable).

In contrast, the transverse (compared to the quantization axis) correlation $\langle S^\pm_0(t)S^\mp_0(0)\rangle$ does not display any algebraic decay at long time. This is because $S^z_\mathrm{tot}$ commutes with the Hamiltonian and naturally appears in the momentum space formulation of the longitudinal correlation function $\langle S^z_0(t)S^z_0(0)\rangle = \sum_{q}\langle S^z_{-q}(t)S^z_{q}(0)\rangle/L$ at the $q=0$ point; while it does not in the transverse case. Spin diffusion implies that the NMR relaxation $1/T_1$ explicitly depends on the NMR frequency as $1/T_1\approx \omega_0^{\alpha-1}$ ($\omega_0$ is playing the role of a cutoff). Ultimately, a dominant contribution from $q\approx 0$ modes, especially over $q\approx \pi$ modes which one naively expects to dominates in antifferomagnets, is a signature of spin diffusion~\cite{Sandvik1995}. For instance, this has been observed in the one-dimensional $S=1/2$ ~Sr$_2$CuO$_3$~\cite{thurber2001} and Cu(C$_4$H$_4$N$_2$)(NO$_3$)$_2$~\cite{kuhne2009} compounds, as well as in the $S=1$ compound AgVP$_2$S$_6$~\cite{takigawa1996}.

Using MPS, we are able to compute the time-dependent local spin-spin correlation function of the $S=1$ chain as shown in Fig.~\ref{fig:spin_diffusion}. At high temperature $\beta J\lesssim 0.8$, an algebraic form is observed at long time $t$. The diffusion exponent obtained by fitting the long-time decay by a simple power-law is found to be roughly independent of temperature and in average equal to $\overline{\alpha}\simeq 0.62$. This value is not equal to $1/2$, as it would be expected for a non-integrable model, nor is it equal to $2/3$, as for the isotropic spin-half Heisenberg chain, and as suggested by a recent field-theory approach~\cite{DeNardis2019}. According to Ref.~\onlinecite{dupont2019}, where a more systematic study of the long-time algebraic decay was performed, the value of $\alpha$ that we get is simply a crossover value due to the finite time, and will eventually reach $1/2$ at longer time.

\subsection{Contributions of antiferromagnetic modes ($q\approx \pi$) to spin relaxation}

\begin{figure}[t]
    \includegraphics[width=\linewidth,clip]{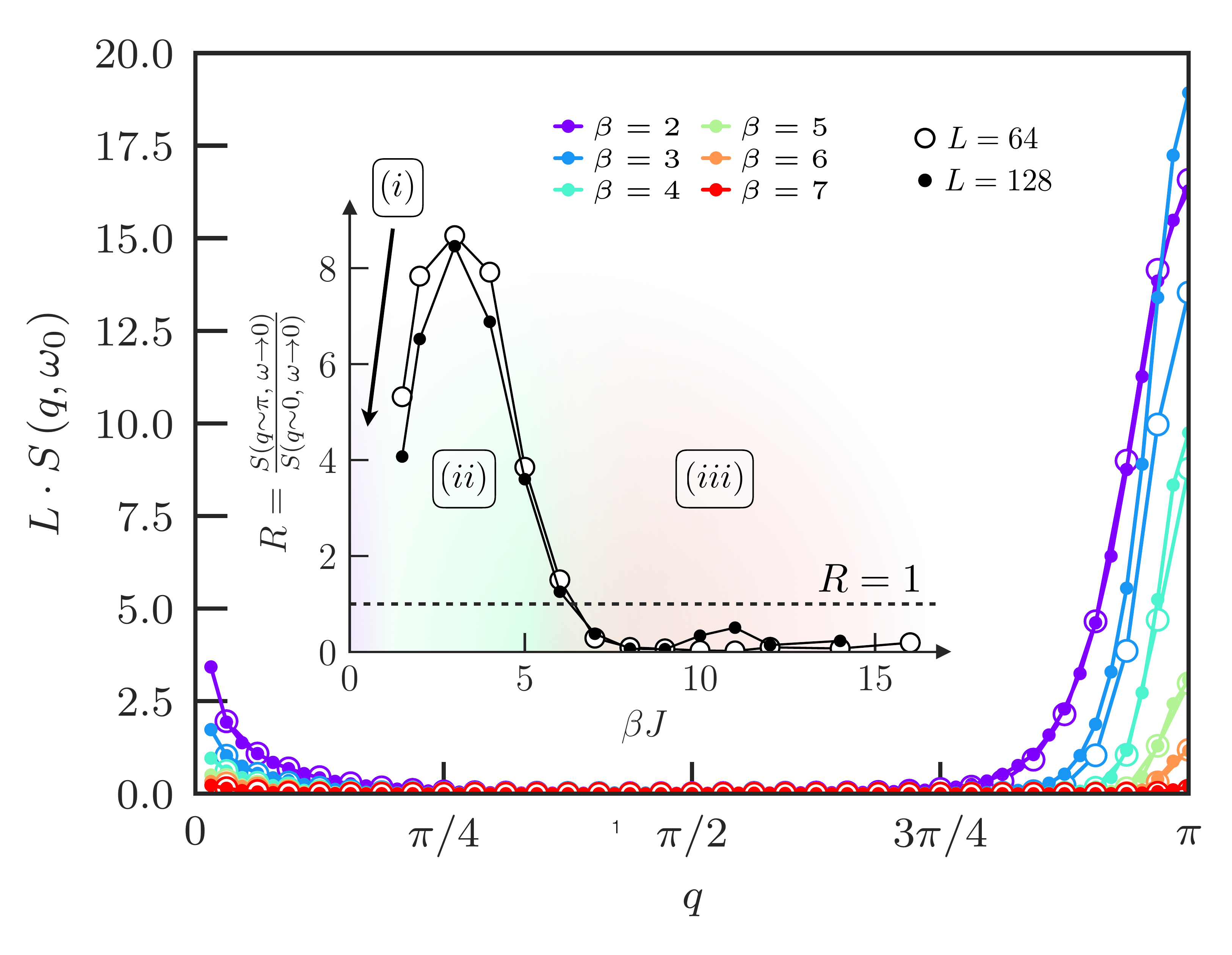}
    \caption{Momentum  contribution $S(q,\omega_0\simeq 0)$ to $1/T_1$ obtained from QMC simulations on chains of length $L=64$ and $L=128$ for various inverse temperatures $\beta J=2,3,4,5,6$ and $7$. The values have been multiplied by $L$ so that we can compare systems with different lengths $L$. The inset shows the ratio $R$ of the contributions close to $q\approx \pi$ versus $q\approx 0$, as defined in Eq.~\eqref{eq:R}. Three different regions are observed: (i) corresponds to spin diffusion at high temperature, dominated by $q\approx 0$ modes with an explicit dependence of the $1/T_1$ value on the NMR frequency $\omega_0$. In region (ii), at inverse temperatures around the inverse spin gap, i.e. $\beta\approx 1/\Delta=2.44$, the spin-lattice relaxation rate is dominated by the $q\approx \pi$ contribution as one might expect from an antiferromagnet. However, as the temperature is lowered, a crossover is observed where $q\approx 0$ contributions start to dominate again in region (iii), see discussions in text.}
    \label{fig:T1q}
\end{figure}

Although we are ultimately interested in the local dynamical spectral function, we focus now on small $\omega$ for the entire $S(q,\omega)$ at finite temperature, in order to investigate which momenta contribute significantly to $1/T_1$. Indeed from the above definition of Eq.~\eqref{eq:T1}, the spin-lattice relaxation rate is simply given as a sum over momenta $q$ of $S(q,\omega_0)$. We plot in Fig.~\ref{fig:T1q} these contributions to $1/T_1$ as a function of momentum $q$. In order to check the finite-size effect, we multiply by $L$ so that $1/T_1$ is proportional to the integral under the curves and we can compare systems with different lengths $L$. As expected, since this is a local dynamical quantity, it does not depend much on $L$ at fixed $\beta$. Regarding the temperature dependence, we propose to quantify the relative importance of momenta close to $0$ and $\pi$ by defining the ratio,
\begin{equation}
    \label{eq:R}
    R = \frac{\sum_{\frac{\pi}{2}\leq |q|\leq \pi}{S(q,\omega_0)}}{\sum_{0\leq  |q|\leq\frac{\pi}{2}}{S(q,\omega_0)}}.
\end{equation}
As seen in Fig.~\ref{fig:T1q}, there is a rather sharp crossover around $\beta J\simeq 6$ below (above) which low-energy spectral weight is mostly at $q\approx \pi$ ($q\approx 0$). It can be expected that, due to antiferromagnetic interaction, excitations with momentum $q\approx \pi$ should be important, and thus $R>1$. However, as has been known for a long time~\cite{jolicoeur1994,sagi1996},  relaxation is in fact dominated by two-magnon processes at very low-temperature (due to energy conservation), hence we do expect $q\approx 0$ to dominate at low enough temperature (much lower than the gap), which is illustrated when $R<1$ at large $\beta$. We can also observe a divergence at small $q$ for any finite temperature $T$, which would correspond to spin diffusion. Indeed, our QMC data can be well fitted as $\propto q^{-1}$ for small $q$ (data not shown), so that we formally get a divergence as $\ln L$ which would correspond to spin diffusion: namely, the $1/T_1$ relaxation rate does depend explicitly on a cutoff, which is experimentally the NMR frequency $\omega_0$. We seem to observe the same behavior for any temperature, with an exponent $\alpha=1$ quite different from our MPS results for which we found $\alpha\simeq 0.62$. However, as stated above, we do not trust quantitatively SAC results at small $\beta$, see Fig.~\ref{fig:compare_beta2}.

\subsection{Temperature dependence of $1/T_1$}

\begin{figure}[t]
    \includegraphics[width=\linewidth,clip]{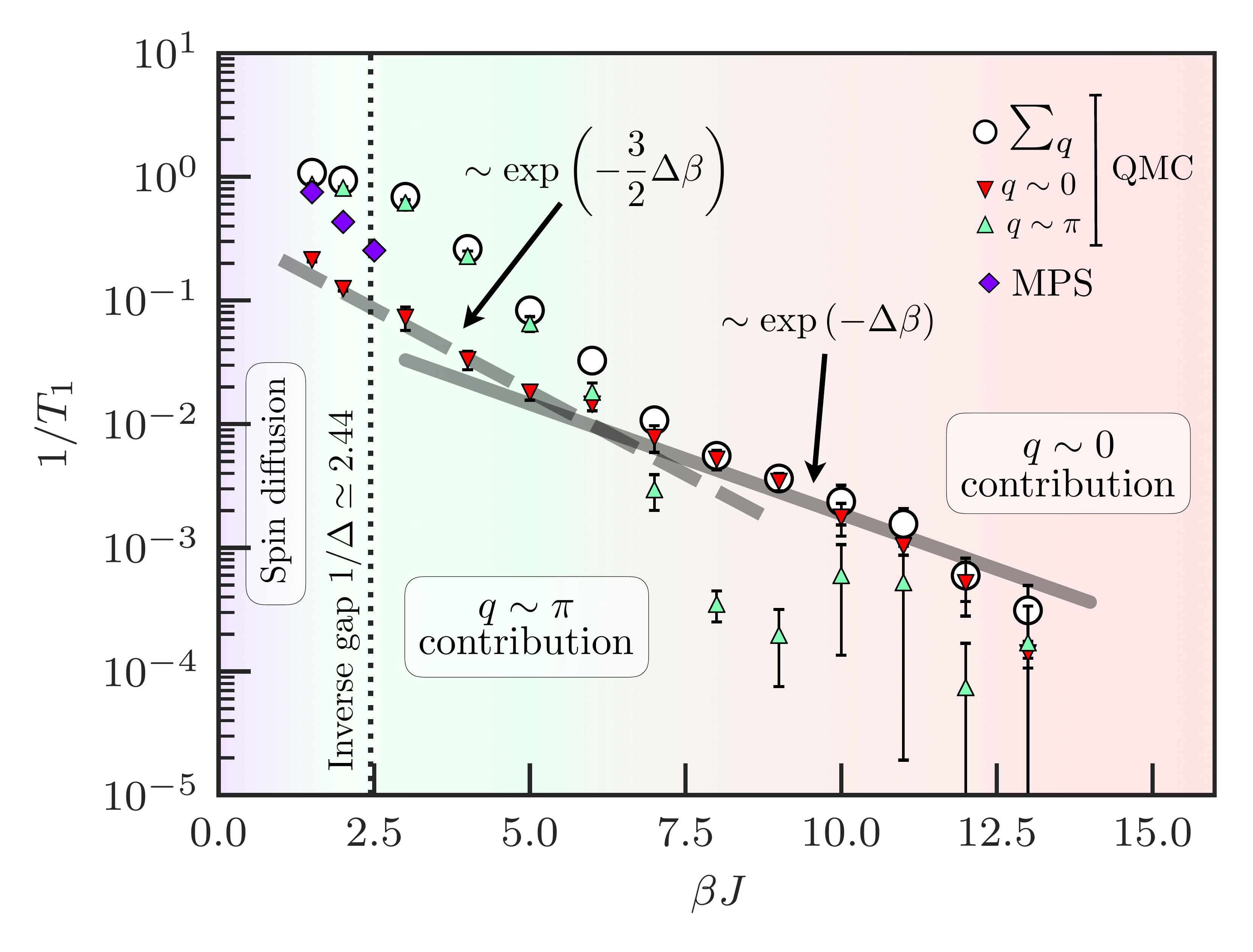}
    \caption{NMR spin-lattice relaxation rate $1/T_1$ versus inverse temperature $\beta$ obtained from QMC simulations on a chain with $L=128$ spins (white circles), and from MPS calculations on a chain with $L=64$ spins. In the latter case, we show only data where $G_\mathrm{loc}(\omega_0)$ could be precisely determined, i.e. with a ``relatively fast'' decay of the correlation $\langle \mathbf{S}_{L/2}(t)\cdot\mathbf{S}_{L/2}(0)\rangle$ to perform a proper Fourier transform despite the maximum time $t$ we could reach in practice. We also plot separately the contributions coming from $q\approx 0$ (downward red triangles) and $q\approx \pi$ (upward green triangles) of the QMC data. The gray solid (dashed) line corresponds to a simple (modified) activated law $\propto\exp(-\beta\Delta)$ ($\propto\exp[-(3/2)\beta\Delta]$). For discussion, we have shown the inverse spin gap $1/\Delta\simeq 2.44$. Spin diffusion data at high temperatures have been discarded since they explicitly depend on the NMR frequency $\omega_0$ and were discussed earlier on their own.}
    \label{fig:T1_beta}
\end{figure}

Combining our numerical results, we can obtain the full behavior of $1/T_1$ versus inverse temperature $\beta$, see Fig.~\ref{fig:T1_beta}. At high temperature, we are more confident in MPS results since imaginary-time simulation are limited to small time $\tau=\beta/2$ and cannot produce very reliable results~\cite{Shu2018}, but for temperature above the Haldane gap ($\beta\lesssim 1/\Delta$), there is a spin diffusion regime where the $1/T_1$ depends explicitly on the cutoff procedure, see above.

For intermediate and low temperature, we can rely on only QMC simulations since real-time data obtained from MPS are limited to time $tJ \approx 50$ and strong oscillations prevent a reliable estimate of the Fourier transform (note that $1/T_1$ becomes exponentially suppressed). As already seen in Fig.~\ref{fig:T1q}, there is another crossover between a regime with dominant $q\approx \pi$ contributions ($1/\Delta \lesssim \beta J \lesssim 6$) where $1/T_1$ decreases very fast. Note however that, in this intermediate temperature regime, the subdominant $q\approx 0$ contributions would be compatible with a modified activated law $\propto\exp[-(3/2)\Delta \beta]$. At lower temperature, the signal becomes extremely small and in order to have some intuition into the quality of our data, we have performed a bootstrap analysis of our QMC data using ten bootstrap samples followed by SAC. We extract some tentative error bars from this analysis, see Fig.~\ref{fig:T1_beta}. In this low-temperature regime $\beta J\gtrsim 6$, we observe that $q\approx 0$ contributions are dominant and overall behavior seem to better follow  a simple activated law $\propto\exp(-\Delta \beta)$.

\section{Discussion and outlook}\label{sec:conclusion}

Our numerical study relies on different numerical techniques: MPS, expected to be valid at high-tempeature and SAC on top of QMC simulations which we trust for low-temperature studies. We can summarize our key findings in different temperature regimes:
\begin{enumerate}[label=(\roman*)]
    \item High-temperature ($T> \Delta$): We have observed a spin diffusion regime which prevent any universal prediction since by definition, $1/T_1$ depends explicitly on some numerical cutoff or experimental parameters. More precisely, our MPS data suggest an exponent in real time correlation data which is not really one half, as expected for a non integrable model, but rather $\alpha\simeq 0.62$. According to Ref.~\onlinecite{dupont2019} where a more systematic study was performed, this is a finite time effect, and the value that we get is simply a crossover which will eventually go to $1/2$ at longer time.
    \item Intermediate regime ($1/\Delta<\beta J \lesssim 6$): We have shown that the dominant contributions to $1/T_1$ come from momenta $q \approx\pi$, which can be simply interpreted since there is a large Lorentzian peak at the antiferromagnetic wavevector $\pi$ and single magnon excitation can be thermally excited in this temperature range.
    \item Low-temperature regime ($\beta J \gtrsim 6$): Due to energy conservation, the dominant contribution to $1/T_1$ is due to two-magnon processes~\cite{jolicoeur1994,sagi1996} and occurs at $q\approx 0$. Our best data in this regime are compatible with a simple activated law $1/T_1\propto\exp(-\beta \Delta)$, as shown in Fig.~\ref{fig:T1_beta}.
\end{enumerate}

From an experimental point of view, one has to remember that a precise comparison of $1/T_1$ depends on which nucleus is probed and what the hyperfine couplings to the electronic spins are. Indeed, for instance if the NMR nucleus is coupled symmetrically to two $S=1$ magnetic ions, then the $q\approx \pi$ contributions will be filtered out due to the form factors. We have shown that there is a nontrivial crossover when comparing the $q\approx\pi$ and $q\approx 0$ component contributions as shown in Fig.~\ref{fig:T1_beta}. As a result, we do expect that depending on the NMR details (such as the nucleus probed and hyperfine couplings), the temperature behavior could be non-universal in a temperature range of the order of the spin gap. This could explain the various results obtained when comparing the activation energy from $1/T_1$ and the spin gap (which can also be extracted from the NMR signal through the Knight shift for instance), as discussed in the Introduction when summarizing experimental measurements. Also, it could hinder any attempt to fit the $1/T_1$ behavior in this intermediate regime.

Some puzzles remain however, for instance AgVP$_2$S$_6$ in which the activation gap seems larger than the spin gap, independent of the NMR nucleus~\cite{takigawa1996}, roughly compatible with a $\gamma$ factor of 3/2 as found in some theoretical predictions~\cite{sachdev1997,DeNardis2019}. As pointed out by Konik in Ref.~\onlinecite{konik2003}, one cannot exclude that easy axis spin anisotropy, weak interchain interaction or spin-phonon couplings could qualitatively change the temperature behavior. In particular, these additional ingredients are already known to modify strongly spin diffusion~\cite{Fujimoto1999}. Let us also point out that, since the spin gap is field dependent, the finite magnetic field needs to be taken into account for a quantitative analysis~\cite{Fujiwara1992,Fujiwara1993,Reyes1997,Sato1998}. In some other related systems, such as an explicit dimerized $S=1/2$ chain (which is also a one-dimensional gapped system), it has been shown for instance that a simple activated law $1/T_1\propto \exp(-\beta \Delta)$ holds~\cite{coira2016,coira2018} for the NMR relaxation rate. It will be interesting to investigate other simple, yet non trivial, gapped one-dimensional systems such as spin-$1/2$ ladder or other dimerized chains where a simple activated law is often measured~\cite{azuma1994,ishida1994,furukawa1996,kikuchi1997,iwase1996,Isobe1996,Lue2007}.

As a final remark, it would be interesting to investigate NMR relaxation for a more general spin-1 model, such as the XXZ chain with Ising anisotropy $\Delta$ or single-site anisotropy $D$, which has nontrival spin dynamics at finite temperature~\cite{lange2018,richter2019}, or even more relevant for experiments, a quasi-one-dimensional system~\cite{Wierschem2014}.

\begin{acknowledgments}
    The authors are grateful to Mladen Horvati\'c and Nicolas Laflorencie for valuable discussions. M.D. was supported by the U.S. Department of Energy, Office of Science, Office of Basic Energy Sciences, Materials Sciences and Engineering Division under Contract No. DE-AC02-05-CH11231 through the Scientific Discovery through Advanced Computing (SciDAC) program (KC23DAC Topological and Correlated Matter via Tensor Networks and Quantum Monte Carlo). A.W.S. was supported by the NSF under Grant No.~DMR-1710170 and by a Simons Investigator Award. P.S. was supported by grant MOE2016-T2-1-065 from the Ministry of Education, Singapore. The computations were performed using HPC resources from GENCI (Grant No. x2016050225 and No. x2017050225), CALMIP, and Boston University's Shared Computing Cluster. S.C and M.D. acknowledge support of the French ANR program BOLODISS (Grant No. ANR-14-CE32-0018), R\'egion Midi-Pyr\'en\'ees, the Condensed Matter Theory Visitors Program at Boston University and Programme Investissements d'Avenir within the ANR-11-IDEX-0002-02 program, reference ANR-10-LABX-0037-NEXT.
\end{acknowledgments}

 \include{draft.bbl}

\end{document}

%% file: draft.bbl
%